\documentclass[journal]{IEEEtran}

\usepackage[pdftex]{graphicx}

\usepackage{amsmath}

\usepackage{currfile}
\usepackage{standalone}
\usepackage{tikz}
\usepackage{tikzscale}
\usetikzlibrary{decorations.markings}
\usepackage{pgfplots}
\pgfplotsset{compat=newest,grid=major,major grid style={black,dotted},every
	axis plot/.append style={thick}}
\usepackage{siunitx}
\DeclareSIUnit{\belisotropic}{Bi}
\DeclareSIUnit{\dBi}{\deci\belisotropic}
\DeclareSIUnit{\bit}{bit}

\usepackage{booktabs}

 \usepackage{flushend}
\usepackage{lipsum}

\usepackage[acronym]{glossaries}

\newacronym{DAC}{DAC}{Digital to Analog Converter}
\newacronym{DC}{DC}{Direct Current}
\newacronym{DRIE}{DRIE}{Deep Reactive Ion Etching}
\newacronym{IMSL}{IMSL}{Inverted MicroStrip Line}
\newacronym{ITO}{ITO}{Indium Tin Oxide}
\newacronym{LCD}{LCD}{Liquid Crystal Display}
\newacronym{LOS}{LOS}{Line-Of-Sight}
\newacronym{MEMS}{MEMS}{Micro-Electro-Mechanical Systems}
\newacronym{mm-Wave}{mm-Wave}{millimeter-Wave}
\newacronym{NLC}{LC}{nematic Liquid Crystal}
\newacronym{NLOS}{NLOS}{Non Line-Of-Sight}
\newacronym{PIN}{PIN}{Positive-Intrinsic-Negative}
\newacronym{RF}{RF}{Radio-Frequency}
\newacronym{RIS}{RIS}{Reconfigurable Intelligent Surface}
\newacronym{SNR}{SNR}{Signal-to-Noise Ratio}
\newacronym{TFT}{TFT}{Thin-Film Transistor}
\newacronym{RA}{RA}{Reflect-Array}
\newacronym{PA}{PhA}{Phased-Array}
\newacronym{SC}{SC}{Semiconductor}
\newacronym{NRCS}{NRCS}{Normalized RIS Cross-Section}

\usepackage{enumerate}

\usepackage{xcolor}

\definecolor{DarkGreen}{RGB}{0,150,0}

\definecolor{Orange}{RGB}{245,100,10}

\usepackage[normalem]{ulem}

\usepackage{todonotes}

\begin{document}

\title{Reconfigurable Intelligent Surfaces with Liquid Crystal Technology: A Hardware Design and Communication Perspective}

\author{Alejandro Jim\'enez-S\'aez,  Arash Asadi,  Robin Neuder, Mohamadreza Delbari, and Vahid Jamali
\thanks{A. Jim\'enez-S\'aez and R. Neuder  are with the Institute of Microwave Engineering and Photonics at Technical University of Darmstadt, Darmstadt, Germany (e-mail:  \{alejandro.jimenez\_saez, robin.neuder\}@tu-darmstadt.de).} 
\thanks{A. Asadi is with the Wireless Communication and Sensing Lab at Technical University of Darmstadt, Darmstadt, Germany (e-mail:  aasadi@wise.tu-darmstadt.de).} 
\thanks{M. Delbari and V. Jamali are with the Resilient Communication Systems Lab at Technical University of Darmstadt, Darmstadt, Germany (e-mail:  \{mohamadreza.delbari, vahid.jamali\}@tu-darmstadt.de).} 
}



\maketitle

\begin{abstract}
With the surge of theoretical work investigating Reconfigurable Intelligent Surfaces (RISs) for wireless communication and sensing, there exists an urgent need of hardware solutions for the evaluation of these theoretical results and further advancing the field. 
The most common solutions proposed in the literature are based on varactors, Positive-Intrinsic-Negative (PIN) diodes, and Micro-Electro-Mechanical Systems (MEMS). 
This paper presents the use of Liquid Crystal (LC) technology for the realization of continuously-tunable extremely large millimeter-wave RISs. 
We review the basic physical principles of LC theory, introduce two different realizations of LC-RISs, namely reflect-array and phased-array, and highlight their key properties that have an impact on the system design and RIS reconfiguration strategy.   
Moreover, the LC technology is compared with the competing technologies in terms of feasibility, cost, power consumption, reconfiguration speed, and bandwidth. 
Furthermore, several important open problems for both theoretical and experimental research on LC-RISs are presented.
\end{abstract}

\begin{IEEEkeywords}
Liquid crystal, reconfigurable intelligent surface, intelligent reflective surface, millimeter wave, and 6G.
\end{IEEEkeywords}

%
\IEEEpeerreviewmaketitle

\section{Introduction}
\label{sc:introduction}

In recent years, there has been considerable interest in \glspl{RIS}, particularly within the context of 6G communications.  The potential for significant improvements in communication and sensing capabilities through the use of large, passive, and tunable reflectors has led to the development of theoretical models and algorithms in the wireless communication community \cite{di2020smart,yu2021smart} and subsequent attention from the microwave community towards the practical implementation of \glspl{RIS}~\cite{kazim2020wireless}.

\begin{figure}[t]
	\centering
	\includegraphics[width=0.96\columnwidth]{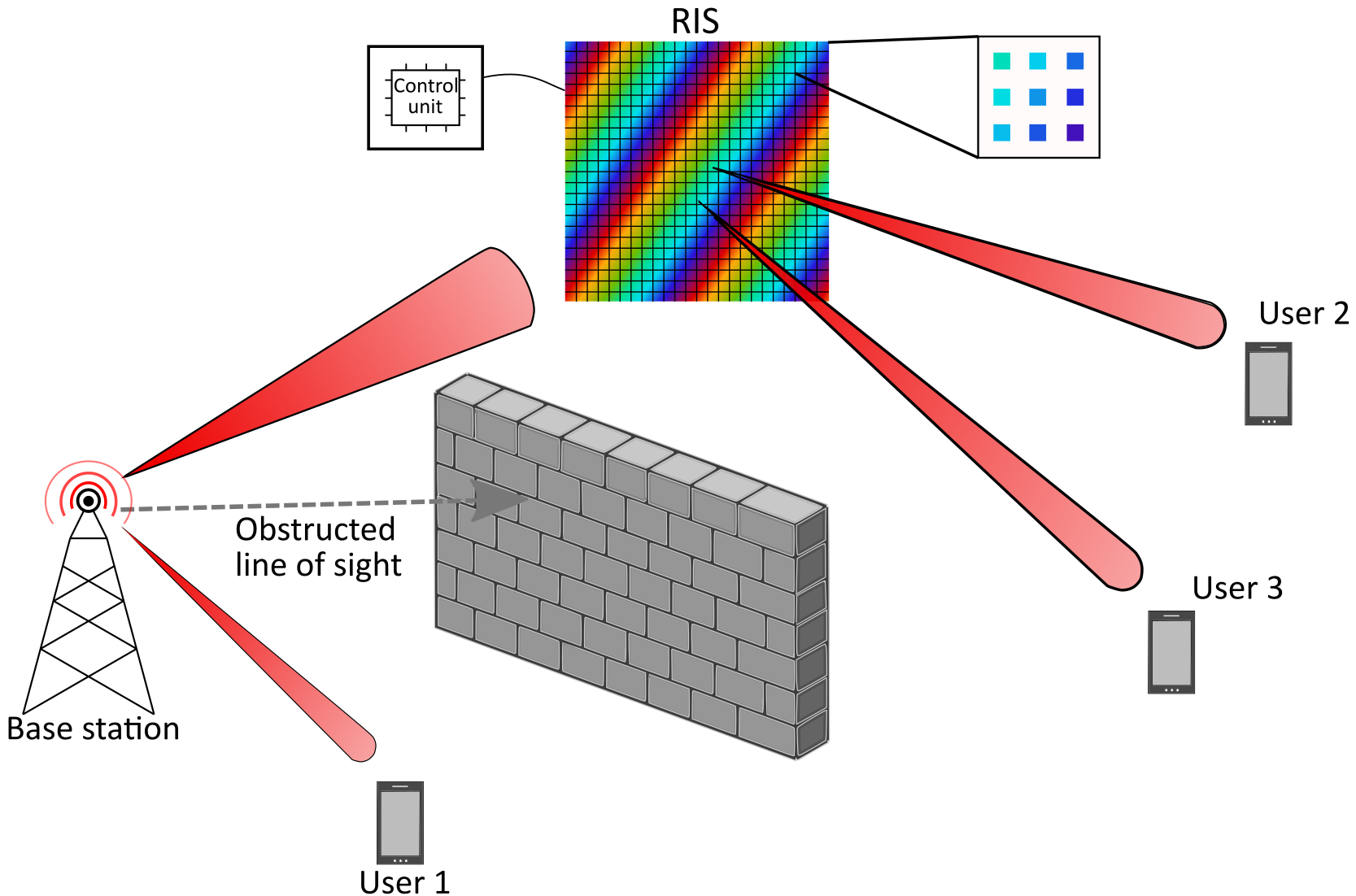}
	\caption{Example scenario where a \gls{RIS} establishes a link between a transmitter and two receivers despite an obstacle blocking the \gls{LOS} path.}
	\label{fig:RISScenario}
\end{figure}

\glspl{RIS} are planar electromagnetic surfaces comprised of many independently tunable reflecting elements. 
These reflecting elements can be adjusted externally to modify the phase and, in certain instances, the amplitude of the signal reflected at each element. 
Through the superposition of reflections from each element, the reflection pattern can be dynamically tuned, eliminating the need for complex decoding, encoding, and processing. 
While demonstrations of reflectarrays and predictions on its tunability date back to 1963 \cite{berry1963reflectarray}, their large-scale application to mobile networks (see Fig.~\ref{fig:RISScenario}) has only been recently considered.

    So far, various \gls{RIS} prototypes have been reported in the literature, which mainly employ \gls{SC} technologies (e.g., \gls{PIN} diodes \cite{tang2020wireless}, varactors \cite{alamzadeh2021reconfigurable}, 
    \gls{RF} switches \cite{rossanese2022designing}) to realize the phase shifts. Although a feasible choice of design, their cost and resolution (in case of \gls{PIN} diodes) can be prohibitive in building truly large and cost-efficient large \glspl{RIS}. An alternative to these technologies is \gls{NLC} technology. In the last two decades, \glspl{NLC} have been studied for the microwave, \gls{mm-Wave}, and THz frequency bands \cite[Ch. 5]{ferrari2022reconfigurable}. 
    If fabrication in standard \gls{LCD} technology is adopted, the manufacturing cost of \gls{NLC}-\glspl{RIS} will be reduced to that of a commercial \gls{LCD} television. Hence, \gls{NLC}-based designs have the advantages of cost-effective scalability, low energy consumption, and continuous phase shifting, which make them a suitable candidate for realizing extremely large passive \glspl{RIS}. Despite these attractive features, \gls{NLC} characteristics impose new challenges when being integrated into today's mobile communication systems, most notably due to their slow tuning capabilities as compared with \gls{SC}-\glspl{RIS}.

In this article, we aim at bridging the gap between microwave and antenna design aspects of \gls{NLC}-\glspl{RIS} with mobile communication aspects. To this end, we first review the basic physical principles of LC theory and introduce two different realizations of \gls{NLC}-\glspl{RIS}, namely the \gls{RA} and \gls{PA} methods. Thereby, we particularly highlight their key physical and hardware properties that have a direct impact on the system design and RIS reconfiguration strategy. This is complemented by a concrete comparison against the alternative technologies for building RISs (i.e., \glspl{SC} and \gls{MEMS})  in terms of feasibility, cost, power consumption, reconfiguration speed, and bandwidth. Finally, we delve into the challenges of \gls{NLC}-\glspl{RIS} (slow response time, temperature dependencies, bandwidth) and present potential directions for both theoretical and experimental research in order to overcome these challenges.

\section{Liquid Crystal-based RISs}

In this section, we first present the basic operating principles of \gls{NLC} technologies. Subsequently, we introduce two of the existing \gls{NLC}-\gls{RIS} implementations, as well as the hardware components required for implementing them.

\subsection{Basic physical principles of LC technology}\label{Sec:Basic}

\begin{figure}[t]
	\centering
 \includegraphics[width=0.8\columnwidth]{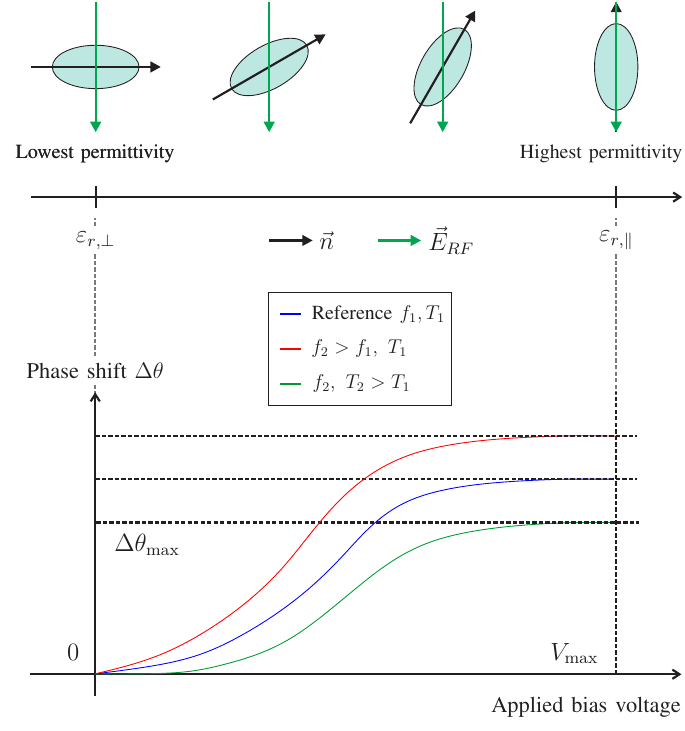}
	\caption{Top figure: Observed \gls{NLC} permittivity depending on the orientation between the \gls{RF} electric field, $\vec{E}_{RF}$, and the \gls{NLC} molecule major axis, $\vec{n}$. The lowest permittivity, $\varepsilon_{r,\perp}$, is observed when $\vec{E}_{RF}$ and $\vec{n}$ are orthogonal while the highest permittivity, $\varepsilon_{r,\parallel}$, is observed when $\vec{E}_{RF}$ and $\vec{n}$ are parallel to each other. Bottom figure: Schematic illustration of the corresponding phase shift achieved by applying bias voltage leading to the change in permittivities. This figure illustrates that the achievable phase shift depends on the operating frequency $f$ and temperature $T$.}
 		\label{fig:LCmolres}
\end{figure}

\textbf{Phase shifting capability:}  
The working principle of \gls{NLC} is based on its electromagnetic anisotropy\footnote{\gls{NLC} molecules exhibit different electromagnetic properties depending on their relative orientation with the \gls{RF} electric field.}. In particular, due to the ellipsoidal shape of \gls{NLC} molecules, the \gls{NLC} presents a larger permittivity (i.e., higher phase shift) when the electric field, $\vec{E}_{\rm RF}$, is aligned with the major axis of the molecules, $\vec{n}$, than when it is aligned with  the minor axis (i.e., lower phase shift), see~Fig.~\ref{fig:LCmolres}. Therefore, by controlling the orientation of \gls{NLC} molecules, we can alter the phase shift of the \gls{RF} signals. 



This is achieved by placing a thin \gls{NLC}-layer between two electrodes. Thereby, when no voltage is applied, the molecules are in their relaxed phase, where $\vec{E}_{\rm RF}$ is perpendicular to $\vec{n}$, and hence \gls{NLC} shows a minimum permittivity $\varepsilon_{r,\perp}$. In contrast, when the maximum voltage is applied, the \gls{NLC} molecules orient along the induced external electric field, which leads to $\vec{E}_{\rm RF}$ being now parallel to $\vec{n}$ and hence a maximum permittivity of $\varepsilon_{r,\parallel}$. The maximum achievable phase shift, $\Delta\theta_{\max}$, therefore scales with $\Delta\varepsilon = {\varepsilon_{r, \parallel}} - {\varepsilon_{r, \perp}}$.  The \gls{NLC}-layer permittivity and the corresponding maximum achievable phase-shift are dependent on the operating frequency as well as temperature, cf. Fig.~\ref{fig:LCmolres}, which will be elaborated further in Section~\ref{sc:challenges}. 

\textbf{Response time:} The alignment of the \gls{NLC} along the electrodes when applying a voltage (for positive phase shifts) is faster than the alignment of the molecules due to mechanical anchoring forces (for negative phase shifts). For this reason, usually the latter slow transition is considered and modelled by the decay or switch-off response time, $\tau_{\rm off}$, which is proportional to $\tau_{\rm off} \propto {\gamma_{\rm rot}{d_{\rm LC}^2}}/{K_{11}}$,  where  $\gamma_{\rm rot}$ is the rotational viscosity, $d_{\rm LC}$ is the \gls{NLC}-layer thickness, and $K_{11}$ is the splay deformation factor of the \gls{NLC}. Therefore, the response time can be reduced by adopting a narrower \gls{NLC} layer; however, this implies more delicate (hence costly) manufacturing and higher conductor losses, motivating research to reduce losses in thin phase shifters \cite{neuder2023compact}.  The \gls{NLC} thickness, $d_{\rm LC}$, can be as low as a few \SI{}{\micro\meter} for switch-off response times, $\tau_\mathrm{\rm off}$, in the order of tens of milliseconds. 

\begin{figure*}[t]
	\centering
\includegraphics[width=1.9\columnwidth]{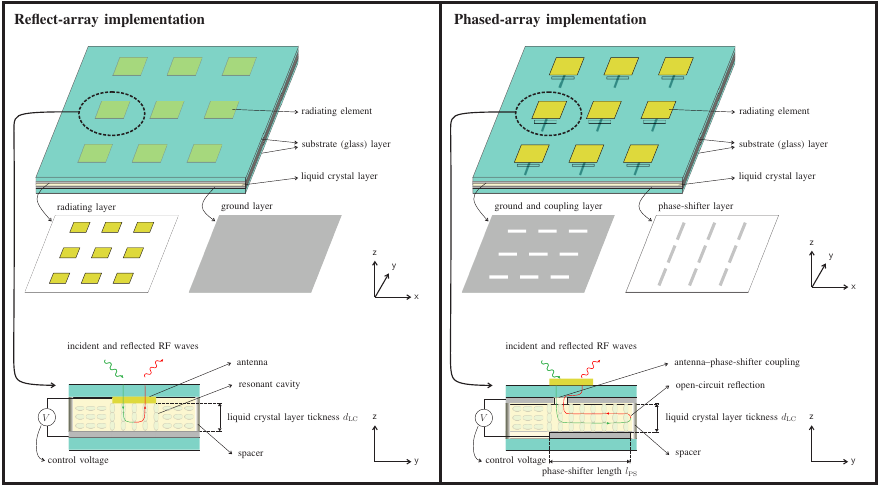}
    \caption{Schematic illustration of reflect-array vs. phased-array LC-RISs. The top figures illustrate the implementation of LC-RISs in three dimensions whereas the bottom figures show a single unit cell from $\mathsf{y}-\mathsf{z}$ cross-section. The different layers, propagation path of the RF wave, and important design parameters such as phase-shifter length $l_{\rm PS}$ (influencing the maximum phase-shift and insertion loss) and LC-layer thickness $d_{\rm LC}$ (impacting the RIS response time) are schematically illustrated.}
    \label{fig:LCrisPARA}
\end{figure*}

\textbf{Insertion loss:} The interaction of the \gls{RF} wave with the \gls{RIS} circuit introduces a certain insertion loss, $L_{\rm ins}$, which depends on the design as well as material properties such as dielectric losses, $\tan\delta$, and conductor losses. For the \gls{PA} implementation (see Section~\ref{sec:RISimplementation}), the phase-shifter length $l_{\rm PS}$ has a direct impact on the final insertion loss and maximum phase shift, see Fig.~\ref{fig:LCrisPARA}. In fact, increasing the phase-shifter length yields a higher maximum phase shift $\Delta\theta_{\max}\propto l_{\rm PS}$ at the cost of a larger insertion loss $L_{\rm ins} \propto l_{\rm PS}$. 

\textbf{Clearing point temperature:} Another physical property of \glspl{NLC} is the temperature at which an \gls{NLC} phase is converted to an isotropic liquid.  This temperature is known as the clearing point, $T_c$, after which tunability is no longer possible, and hence constitutes the highest operation temperatures. 
From the material design perspective, the main targets of \gls{NLC} design is to achieve high phase-shift contrasts, $\Delta\varepsilon $, low dielectric losses, $\tan\delta$, high clearing point temperatures, $T_c$,  and fast response times by reducing the rotational viscosity of the molecules, $\gamma_{\rm rot}$. 
The values of these parameters for several commercially available \gls{NLC} mixtures are shown in Tab.~\ref{tab:LCtypes}.

\begin{table*}
\centering
\caption{LCs for mm and sub-mm wave bands and their material properties. \cite{jakoby2020microwave}}
\label{tab:LCtypes}
\begin{tabular}{@{}lcccccccc@{}} \toprule
LC        & $\varepsilon_{r,\perp}$ & $\tan\delta_\perp$ & $\varepsilon_{r,\parallel}$ & $\tan\delta_\parallel$ & {$\Delta \varepsilon$} & $T_c (^\circ \text{C})$&  $K_{11} (\text{pN})$  &  $\gamma_{\rm rot} (\text{Pa} \cdot \text{s})$ \\ \midrule 
K15 (5CB) & 2.7                     & 0.0273            & 3.1                         & 0.0132                & 0.4                  & 38.0            &    7.0          & 0.126                          \\
GT3-23001 & 2.41                    & 0.0141            & 3.18                        & 0.0037                & 0.77                 & 173.5           &    24           & 0.727                          \\
GT5-28004 & 2.40                    & 0.0043            & 3.32                        & 0.0014                & 0.92                 & 151.0           &    11.8         & 5.953                          \\
GT7-29001 & 2.46                    & 0.0116            & 3.53                        & 0.0064                & 1.07                 & 124.0           &    14.5         & 0.307                        \\ \bottomrule
\end{tabular}
\end{table*}

\subsection{LC-based RIS implementation} \label{sec:RISimplementation}

Next, we first discuss the basic components that are required to realize an \gls{NLC}-\gls{RIS}. Subsequently, we introduce two implementation approaches of \gls{NLC}-\glspl{RIS}, namely the \gls{RA} and \gls{PA} methods, and discuss their advantages and limitations. Both implementations are potentially compatible with current \gls{LCD} technology, which is a great advantage for the realization of cost-effective large \glspl{RIS}, see Fig.~\ref{fig:LCrisPARA}.

\textbf{Basic components:}  The thin \gls{NLC} layer is usually placed between two layers of glass. 
The choice of glass over more common materials is due to the fact that a very thin \gls{NLC} layer with a well-defined thickness needs to be achieved over large panels, therefore a stiff, smooth solid substrate is needed.
Due to its liquid state, solid spacers are needed to maintain the desired \gls{NLC} layer thickness. 
Moreover, the circuits of planar \gls{NLC} are commonly grown directly on glass substrates. Thereby, one or both sides of each glass are selectively metalized to pattern the desired circuitry, e.g., control lines and biasing electrodes. In particular, one of the glass substrates is often metalized to act as a ground layer for biasing all unit cells, whereas the other glass is selectively metalized for introducing phase-shifting control voltage of individual unit cells, see Fig. \ref{fig:LCrisPARA}. 
To receive and re-radiate the phase-shifted signal, a radiating element is needed, usually a patch or a dipole antenna.


\textbf{Reflect-array implementation:} In this method, there is a common ground for all \gls{RIS} elements and the \gls{NLC} layer is directly between the antenna (acting also as one of the biasing electrodes) and the ground, see Fig. \ref{fig:LCrisPARA}. This structure forms a resonant cavity between the metal patch and the uniform ground plane. The advantage of this design is its simplicity since no patterning of the ground plane is necessary and therefore only the layer with the patches must be etched. This avoids any alignment issues during assembly. However, the limitations of this method are: \textit{i)} 
Due to the resonant effect, there is a limited bandwidth at which a nearly constant group delay and low amplitude
variations can be achieved. 
\textit{ii)} Thicker LC layers, $d_{\rm LC}$, are needed to support wideband
resonances with a low radiation quality factor to minimize
the losses. This results in slow response times, {$\tau_{\rm off}\geq 10$s}.
\textit{iii)} The LC biasing lines are comprised in the same layer of the patches and therefore need to be considered in its design.
As a guiding value, such designs achieve impedance bandwidths\footnote{The range of frequencies over which the antenna has an acceptable (-10 dB) impedance matching.} below 10\% and insertion losses in the range 6 to 10 dB.

\textbf{Phased-array implementation:} In this method, one of the glass substrates is metalized such that each \gls{RIS} element has a dedicated phase shifter (i.e., a biasing electrode). Moreover, the ground layer is patterned to couple the radiating elements to the phase-shifters, see Fig. \ref{fig:LCrisPARA}. For the realization of the phase shifters, transmission line topologies compatible with \gls{LCD} manufacturing are needed.
One of these transmission lines is the \gls{IMSL} which is depicted in Fig.~\ref{fig:LCrisPARA}. 
The transmission line should be terminated with a reflective end. 
A suitable solution is the abrupt termination of the metal strip in an open end. 
Although such an open-end shows increased undesired radiation, this radiation is mainly proportional to the thickness of the \gls{NLC} layer, $d_{\rm LC}$. 
Small $d_{\rm LC} <\ $\SI{100}{\micro\meter} hence minimizes this effect.
The advantages of the \gls{PA} method are: \textit{i)} bandwidth is not limited by the phase shifter, and a thick glass in the radiating layer allows a wideband radiating element. A nearly constant group delay and low amplitude variations across the bandwidth can be achieved. 
\textit{ii)} Thin \gls{NLC} layers, $d_{\rm LC}$, are possible and lead to response times, $\tau_{\rm off}<100$~ms.
\textit{iii)} The LC biasing lines are comprised in an additional layer. They do not disturb radiation, but additional processing steps are needed to metalize and pattern the additional layer. The insertion loss of the phase shifter for thin $d_{\rm LC}$, specifically \SI{4.6}{\micro\meter}, is determined to be \SI{4.5}{\dB} for full \SI{360}{\degree} tunability at \SI{28}{\GHz} in~\cite{neuder2023compact}. The matched impedance bandwidth is expected to range around \SI{15}{\percent}. 

\section{Comparison of Technologies for Building \glspl{RIS}}
From a broad perspective, there are three main technologies for realizing a \gls{RIS}, namely \gls{NLC}, \gls{SC}, and \gls{MEMS}. In this section, we provide an overview of each method and discuss the respective advantages and limitations.

\begin{table*}
\centering
\caption{Comparison of Technologies used for the Realization of RIS.}\label{tab:comparison}
\begin{tabular}{llcccc}
\hline
                  & Varactor                                 & PIN Diode                                & \gls{MEMS} Switch                 & \gls{MEMS} Mirror                  & LC                                                            \\ \hline
Tuning time       & ns                                       & ns                                       & $\mathrm{\mu s}$                                        &$\mathrm{100s \: of \: \mu s}$                                     & $>$ 10 ms                                                      \\
Tunability        & continuous                               & discrete (1 bit/diode)                   & discrete                                       & discrete                                            & continuous                                                    \\
Power consumption & medium                                   & medium                                   & low                                            & low                                                 & low                                                           \\
Scalability       & low (cost per diode)                     & low (cost per diode)                     & medium (encapsulation)                          & medium (cost per area)                              & high                                                          \\
Frequency range   & few GHz                                  & mm-Wave                                  & mm-Wave                                        & THz                                                 & $>$ 10 GHz                                                    \\
Cost              & high                                     & high                                     & high                                           & high                                                & low                                                           \\
Example           & \cite{tawk2012varactor} & \cite{tang2020wireless} & \cite{ferrari2022reconfigurable}, Ch. 4 & \cite{schmitt20213} & \cite{neuder2023compact, karabey20122} \\ \hline
\end{tabular}
\end{table*}

\subsection{\gls{NLC}-based \glspl{RIS}}
The fundamentals of \gls{NLC}-\glspl{RIS} have been discussed in detail in the previous section, hence we elaborate on the advantages and limitations of this technology. 

{\bf Advantages:} 
One of the key merits of \gls{NLC}-\glspl{RIS} is {\it scalability at low-cost}. \gls{NLC} has been used for standard \gls{LCD} fabrication processes for decades. 
Therefore, the production of large \gls{NLC} panels is technologically both very accessible and inexpensive. 
The cost of \gls{NLC}-\glspl{RIS} can be even less than \gls{LCD} displays due to the lower number of \textit{pixels}, the lack of backlight, as well as the larger tolerances compared to, for example, the strict color accuracy standards applied to displays. 
The other advantage of \gls{NLC} is its {\it low power consumption}.
Due to its dielectric nature, \gls{NLC} experiences minimal current flow only to alter its state, specifically for rotating the LC molecules. Nonetheless, this power requirement is significantly lower compared to other technologies such as \gls{PIN} diodes. 
The other important feature of \gls{NLC} is the {\it continuous tuning} capability, which is useful when sophisticated wavefront shaping (beyond a simple narrow reflection beam) is required, e.g., to reduce interference in multi-user systems.

{\bf Limitations:} Despite major advances in \gls{NLC} microwave community, the response time of these devices are still higher than \gls{SC} and \gls{MEMS} (e.g., $>10$~ms for \gls{PA} and {$>10$~s} for \gls{RA}). This implies that while the \glspl{RIS} are able to adapt to the user movement (e.g., on the order of seconds), they cannot adapt to fast fading. In addition, \gls{NLC}-\glspl{RIS} are suitable mainly for high frequencies $>10\ $GHz and cannot be adopted for  sub-$6$~GHz communication systems featuring rich multi-path environments. Moreover, the phase-shifting behavior of \gls{NLC}-\glspl{RIS} is temperature-dependent, see Fig.~\ref{fig:LCmolres}, which necessitates the adaption of reconfiguration strategy for the environments with significant temperature variations, see Section~\ref{sec:temperature} for possible solutions. 

\subsection{\gls{SC}-based \gls{RIS}}
\glspl{SC} are very common for \gls{PA} antennas. Similar  designs have been leveraged in prototyping \glspl{RIS}. Under this category, there are two common approaches, namely \gls{PIN} diodes and varactors. 

A \gls{PIN} diode is a low-capacitance device with high-frequency switching capabilities. By toggling between low and high-resistance states, the reflected wave's phase can be switched between two discrete states, typically \SI{180}{\degree} apart. Conceptually, \gls{RF}-switches are similar to \gls{PIN} diodes, as they are most commonly built as a packaged network of \glspl{PIN}. Varactors, on the other hand, have more significant capacitance variations and can be continuously tuned, but this higher capacitance also limits their maximum operation frequency.

{\bf Advantages:} \glspl{SC} are readily available at increasingly higher frequencies. Furthermore, \gls{SC}-\glspl{RIS} offer low insertion loss, fast switching speeds below a \SI{}{\micro\second}, and compact size.

{\bf Limitations:} The high power consumption (varactors, several PIN diodes) and pronounced temperature sensitivity are some of the technical disadvantages to be considered. 
However, the main factor is that the cost of building a \gls{SC}-\gls{RIS} rises quadratically with the surface area due to the increasing number of diodes required (one diode per bit and radiating element). 
Despite continuous advances in packaging, reliably and cost-effectively integrating thousands or even millions of discrete \gls{RF}-components in large surfaces still poses a challenge.
This remains an issue as long as the components cannot be selectively grown in the wafer, as is the case of the \glspl{TFT} used for \gls{NLC} biasing, see biasing challenges in Section~\ref{ss:biasing}.
The main limitation of using the diodes or transistors as the \gls{RF} tuning element is that they need to operate at the \gls{RIS} operating frequency, 
which poses an extremely high demand on the processing capabilities.

\subsection{MEMS-based \gls{RIS}}

From a high-level perspective, \gls{MEMS} phase shifters are structures whose electrically controlled micro-displacements result in phase shifts of the \gls{RF}-fields propagating in the component. These solutions become relevant for high-frequency systems (\gls{mm-Wave} and THz), where the wavelength is small and micro-displacements can lead to significant phase shifts. There are two different methods to use \gls{MEMS} as \gls{RF}-phase shifters: \textit{i})
\gls{MEMS}-actuated mirrors and \gls{MEMS} switches. \glspl{MEMS}-actuated mirrors that change their position in the direction normal to the surface of the antenna, varying the phase of the reflected wave; and \textit{ii)} \gls{MEMS} switches which are in principle  tunable capacitors that rely on the controlled displacement of a conductive structure inside a transmission line, often referred to as a cantilever, with an applied voltage.

\textbf{Advantages:} \gls{MEMS}-actuated mirrors have nearly negligible loss and are faster than \gls{NLC}, in the hundreds of \si{\micro\second} range. \gls{MEMS} switches share the advantage of faster tuning than \gls{NLC}.

\textbf{Limitations:} For \gls{MEMS}-actuated mirrors, the currently achievable maximum micro-displacement of e.g. \SI{150}{\micro\meter} limit the minimum frequency \SI{1}{\tera\hertz} when $360^\circ$ phase shift is desired. 
Furthermore, their element dimension is currently larger than wavelength, thus the presence of grating lobes cannot be avoided.
To overcome the instabilities of the displacements due to the involved non-linear forces, anchoring positions are usually used, which leads to discrete displacements in practical designs.
Nevertheless, multiple-bit solutions exist such as \cite{schmitt20213} with 27 states (more than 4 bits). Finally, the fabrication and packaging is affected by the same cost limitation as \glspl{SC}. 
Similar to actuated mirrors, the stability of \gls{MEMS} switches limits the number of phase-shift states. In addition, the insertion loss of \gls{MEMS} switches is in a similar range as \glspl{NLC} for the lower \gls{mm-Wave} frequencies and increases with frequency.

A comparative summary of the advantages and limitations of the discussed \gls{RIS} technologies are presented in Tab~\ref{tab:comparison}.

\section{Challenges and Future Research}
\label{sc:challenges}
In this section, we discuss some of the unique challenges of \gls{NLC}-based RISs from both system and hardware perspectives.

\subsection{Applications with slow reconfiguration}

One of the main limitations of \gls{NLC}-\glspl{RIS} is their slow response time.  There are two key factors that impact the response time: the \gls{NLC} mixture and the \gls{NLC} layer thickness, i.e., $\tau_{\rm off} \propto \gamma_{\rm rot} d_{\rm LC}^2$.
The rotational viscosity of the \gls{NLC}, $\gamma_{\rm rot}$, can be improved by using less viscous \gls{NLC} mixtures such as GT7 instead of GT5, but at the expense of increased dielectric losses. 
The \gls{NLC} thickness, $d_{\rm LC}$, strongly depends on the adopted \gls{RIS} implementation approach. 
In \glspl{RA}, \gls{NLC} thickness is commonly on the order of \SI{100}{\micro\meter} leading to a response time on the order of tens of seconds. 
Due to the resonant structure in this approach, thinner layers would lead to a significant reduction in both the bandwidth, and an increase in the insertion loss. 
On the other hand, \gls{NLC} thickness in \glspl{PA} is thin as \SI{4.6}{\micro\meter} which yields a response time in the range of tens of milliseconds (e.g., \SI{70}{\milli\second} in \cite{neuder2023compact}). 


The above discussion suggests that \gls{RIS} technology should be chosen based on the applications. For scenarios where the \gls{RIS} configuration is static for an extended time period (e.g. for illumination of a blocked area), the \gls{NLC}-\glspl{RIS} with \gls{RA} implementation are suitable due to relatively lower complexity and cost. For scenarios where continuous adaptation is needed (e.g., for illuminating mobile users/devices), then \gls{NLC}-\glspl{RIS} with \gls{PA} implementation may be a preferred choice. For extremely fast reconfigurations (e.g., to track small-scale fading), other technologies such as \gls{SC}-\glspl{RIS} are needed. Although for the latter case, the overhead of channel estimation and control signaling (rather than \gls{RIS} response time) may be the bottleneck, in practice.


\subsection{Undesired transient behaviour}

\begin{figure*}[h]
     \begin{minipage}[t]{0.205\textwidth}
     \centering
         \includegraphics[width=\textwidth,height=3.5cm]{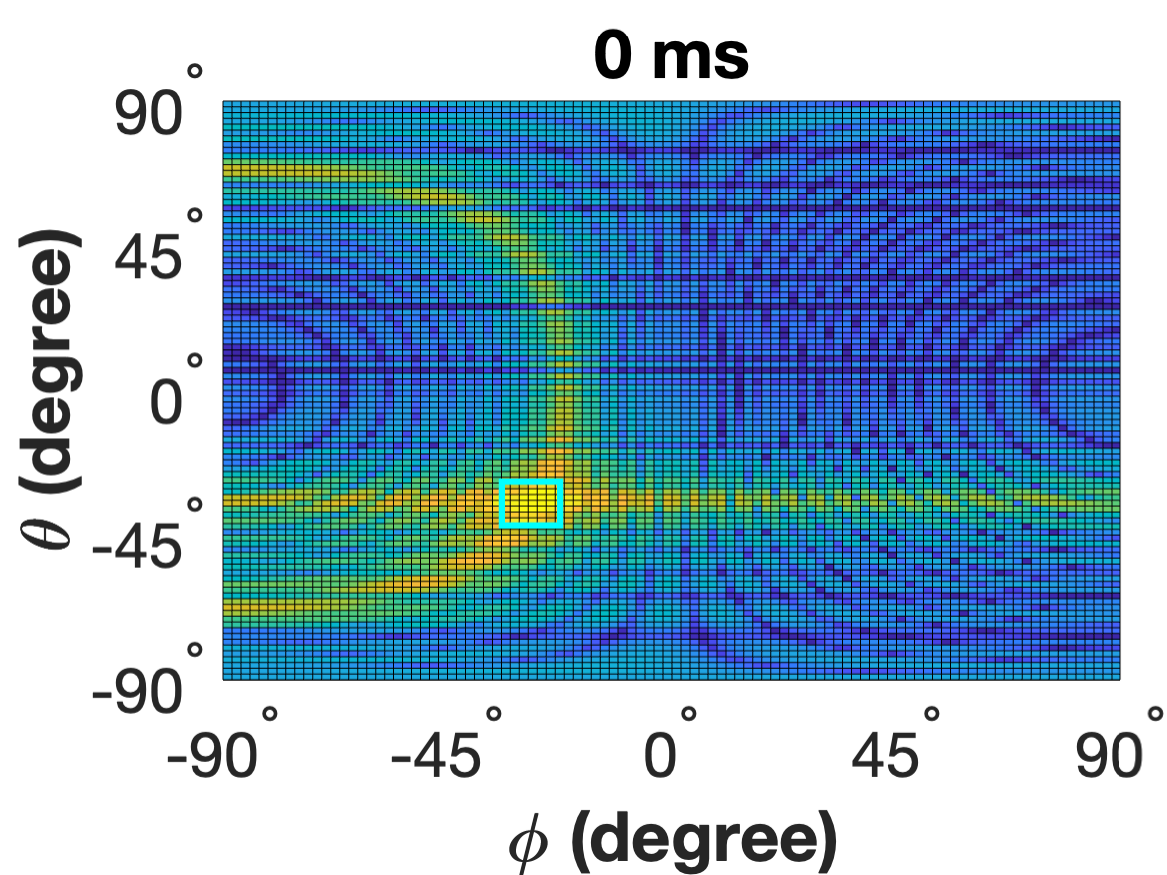}
   \end{minipage}
   \begin{minipage}[t]{0.19\textwidth}
     \centering
         \includegraphics[width=\textwidth,height=3.5cm]{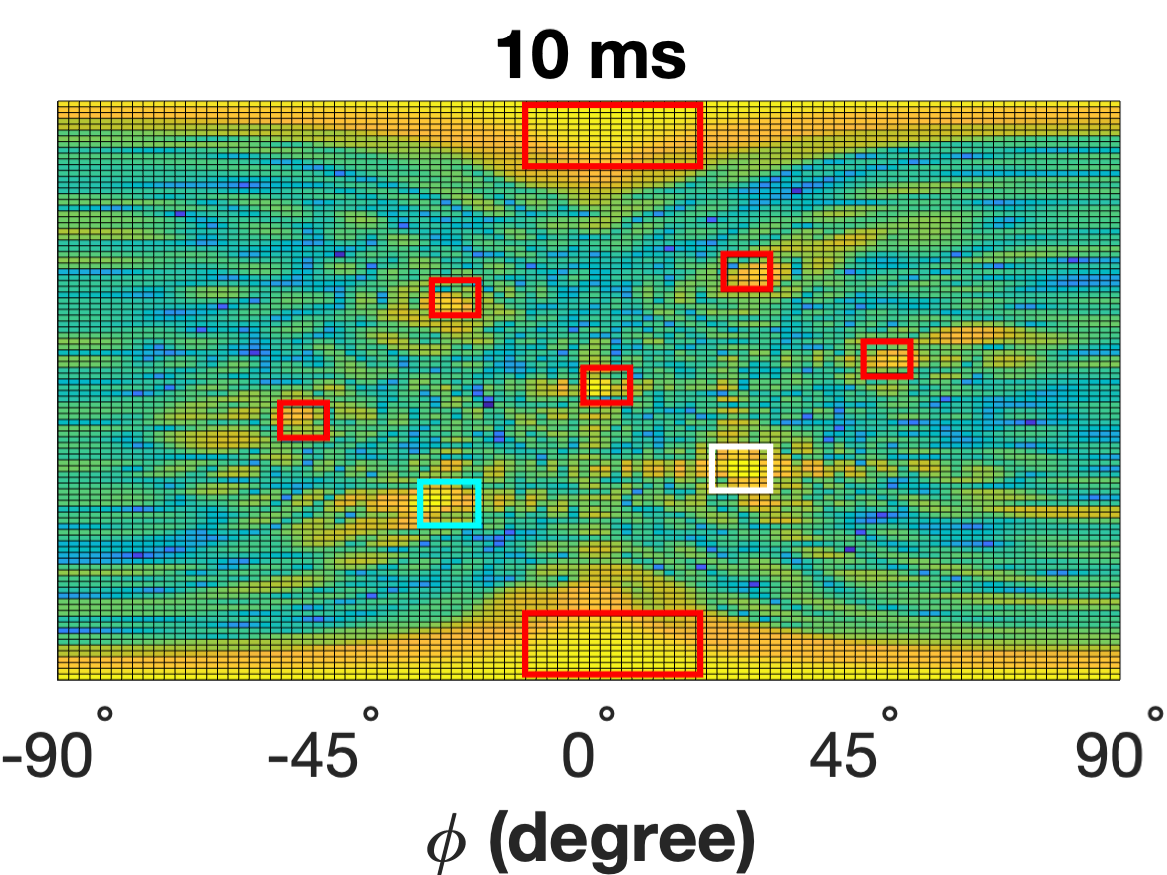}
   \end{minipage}
   \begin{minipage}[t]{0.19\textwidth}
     \centering
         \includegraphics[width=\textwidth,height=3.5cm]{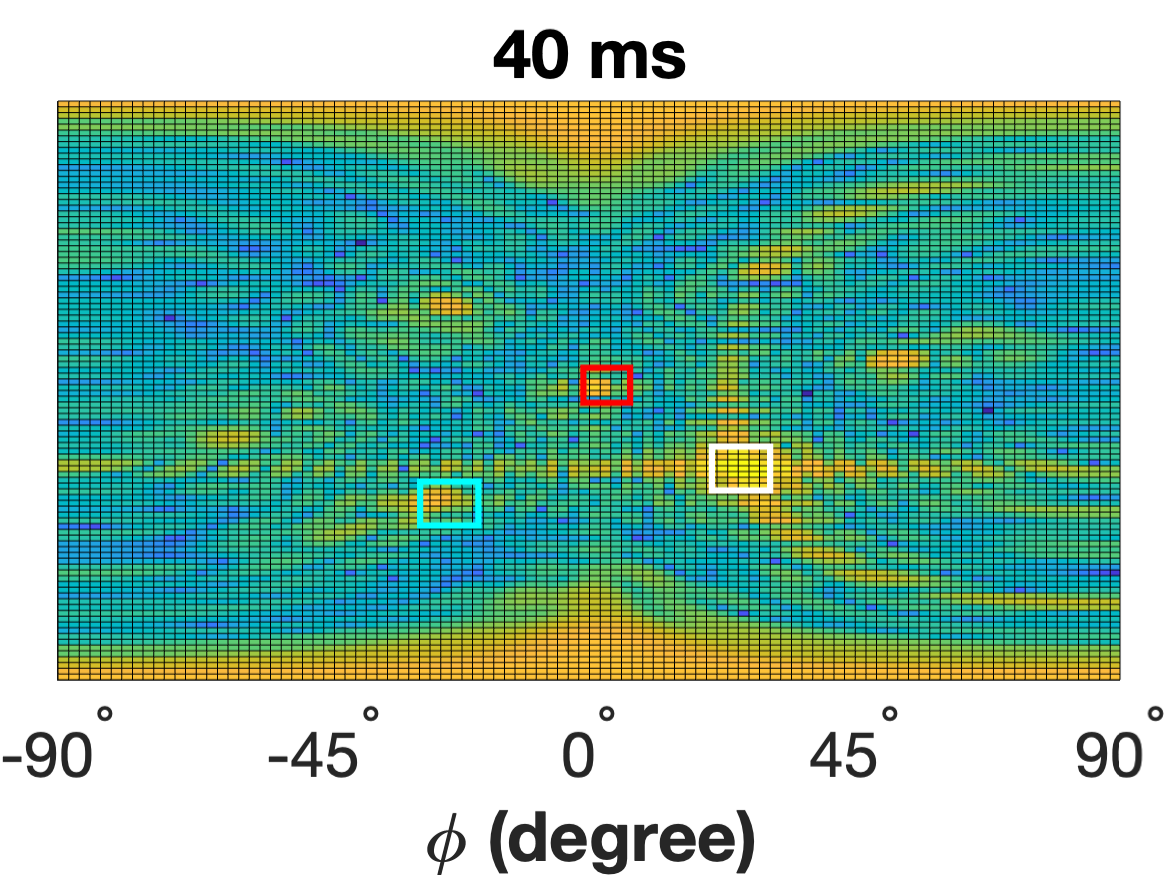}
   \end{minipage}
   \begin{minipage}[t]{0.19\textwidth}
     \centering
         \includegraphics[width=\textwidth,height=3.5cm]{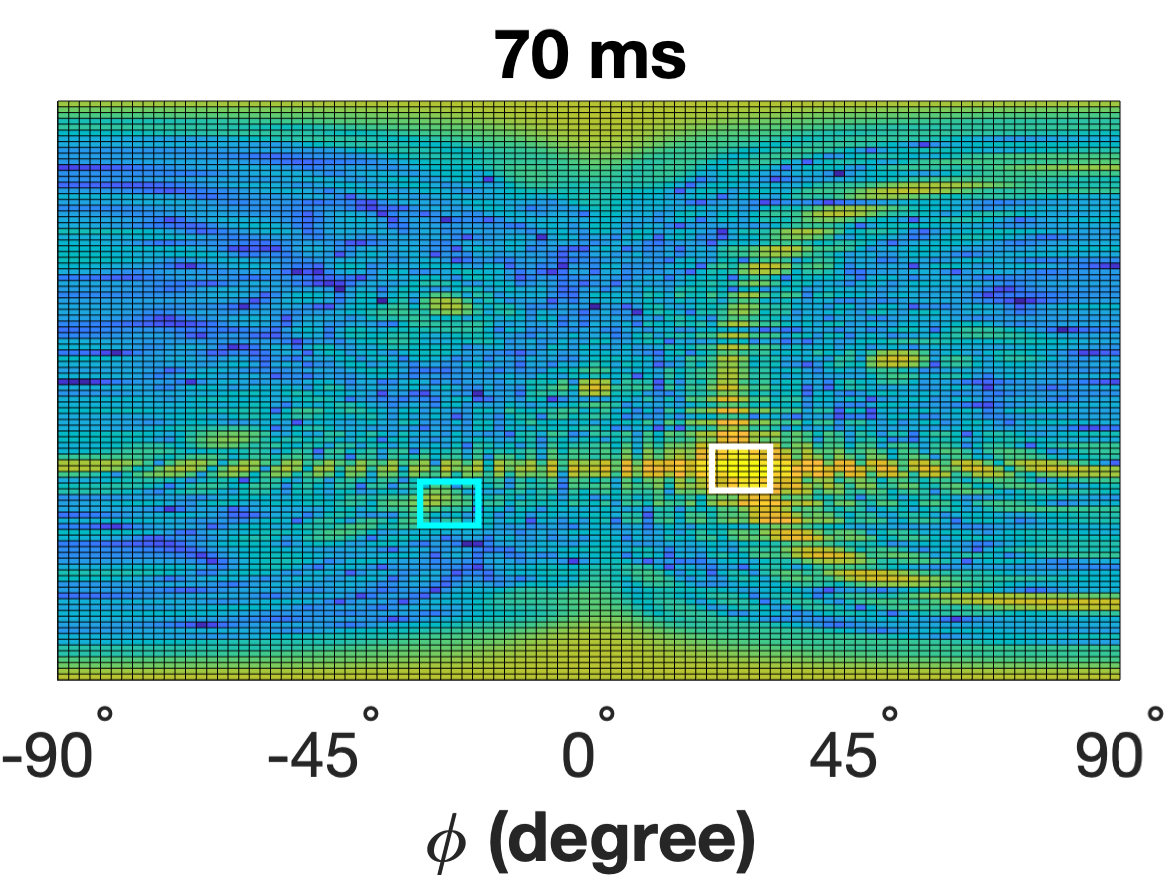}
   \end{minipage}
   \begin{minipage}[t]{0.205\textwidth}
     \centering
         \includegraphics[width=\textwidth,height=3.5cm]{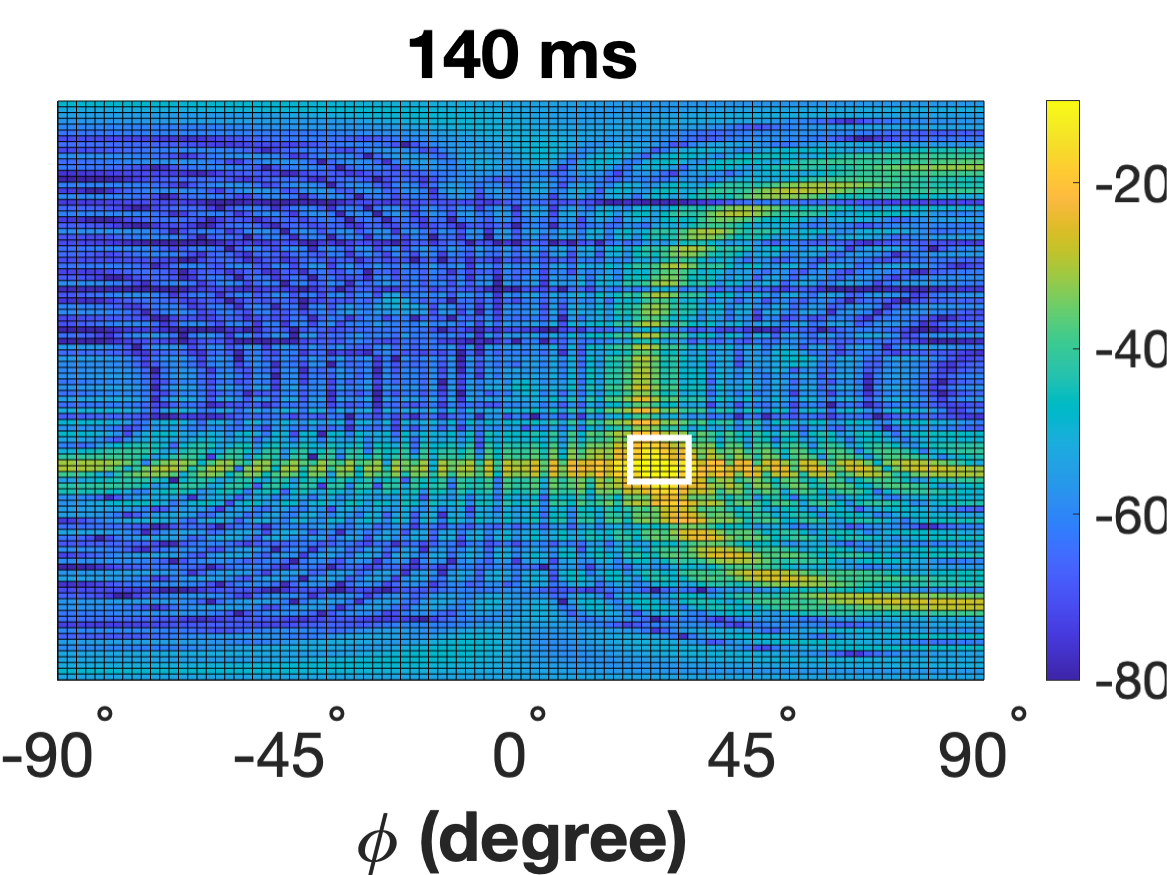}
   \end{minipage}
    \caption{\gls{NRCS} (in dB) at several time instances when transitioning from desired reflection angles $(\phi,\theta)=(-20,-30)$ to $(30,-20)$. Cyan, white, and red rectangles are showing the current, desired, and interference directions, respectively. The \gls{RIS} response time constants are for the \gls{PA} design in \cite{neuder2023compact} and the phase-shifts are based on the quadratic phase-shift design in \cite{jamali2021power}.}
    \label{fig:time response}
\end{figure*}

Due to the slow response time, the reflection pattern of \gls{NLC}-\glspl{RIS} cannot be instantaneously reconfigured, which leads to potentially undesirable transient behavior. We show this transient behavior in Fig.~\ref{fig:time response} for the following case study. It is assumed that an incident wave is normally impinged on the \gls{RIS} and then reflected first at $(\phi,\theta)=(-20,-30)$ and then at $(30,-20)$. For robustness and in order to reduce the reconfiguration overhead, wide beams of approximately 5 degrees are constructed using the quadratic phase shift design from \cite{jamali2021power}. Moreover, we show the \gls{NRCS} where the normalization is w.r.t. the maximum achievable gain by \gls{RIS} (i.e., the peak of \gls{NRCS} is 1 for speculator reflection). The transition behavior is modeled by $\tau_{\rm off}=70$~ms and $\tau_{\rm on}=10$~ms (time where $90\%$ of the phase-shift is achieved), which are taken from the \gls{PA} design in \cite{neuder2023compact}. The \gls{RIS} has a size of $50\lambda\times 50\lambda$, where $\lambda$ is the wavelength. 

Fig.~\ref{fig:time response} shows that the angular profile of \gls{NRCS} (in dB) at time instances $t=0,10,40,70,140$~ms. We can observe from this figure that the \gls{RIS} reflects the wave in many directions within the transient duration, causing interference in unwanted directions. Therefore, an interesting  direction for future research is to develop novel phase-shift configuration strategies that are aware of the \gls{RIS} transient behavior. Example problems include the design of phase-shift schemes that minimize the time to reach the desired reflection pattern or keep the interference during the transient duration below a maximum level.   


\subsection{Temperature-adaptive RIS configuration}\label{sec:temperature}
In an outdoor scenario, a \gls{RIS} needs to withstand a large range of temperatures.
For outdoor cellular systems, operation within \SIrange{-30}{+50}{\celsius} should be guaranteed. 
As is illustrated in Fig.~\ref{fig:LCmolres}, the phase-shift vs. control voltage curve of the \gls{NLC}-\gls{RIS} is a function of temperature. This implies that a \gls{RIS} phase-shift configuration that is designed for temperatures in summer may not be applicable in winter. Hence, there is a need to develop temperature-aware phase-shift strategies. This requires the \gls{RIS} phase-shift behavior to be characterized offline and  the temperature to be estimated (e.g., by a thermometer deployed on the \gls{RIS}) during online operation for adapting the phase-shift design policy. Depending on the estimation error and update frequency of temperature information, robust designs that are valid for a range of temperatures may be required.


\subsection{LC-RIS Design Tradeoffs}

In general, the maximum attainable phase-shift  of \gls{NLC} decreases with increasing temperature or decreasing frequency (see Fig.~\ref{fig:LCmolres}) \cite{tesmer2021temperature}. Therefore, to guarantee the full $360$-degree phase shift tuneability, one should design the \gls{RIS} unit cells for the maximum temperature and minimum operating frequency. This can be achieved in the \gls{PA} implementation by choosing a sufficiently large phase-shifter length $l_{\rm PS}$ (within the feasible range given the implementation constraints), since  the maximum phase-shift scales with $l_{\rm PS}$. However, as discussed in Section~\ref{Sec:Basic}, the insertion loss also scales with $l_{\rm PS}$.
Therefore, investigating the tradeoff among the maximum achievable phase shift, the resulting insertion loss, and the achievable performance (across the operating temperature and frequency) constitutes an interesting direction for future research.
To the authors' knowledge, this effect is not being considered and reported in current \gls{NLC}-\gls{RIS} designs.


\subsection{Biasing}
\label{ss:biasing}
For the realization of large \glspl{RIS}, the use of glass as a substrate is of great advantage due to the compatibility with current \gls{LCD} technology.
As in commercial \gls{LCD} displays, using \glspl{TFT} at each \gls{RIS} element to address them with $N+M$ (number of rows + columns) connecting lines instead of $N \times M$ (one line per element) is a meaningful approach to the challenge of biasing thousands or even millions of elements.
The additional challenge compared to common \glspl{LCD} is the higher capacitance of \gls{mm-Wave} circuits due to their larger size, requiring larger \glspl{TFT}. 
However, constraints might arise to minimize the effect of the bias lines in the \gls{RIS} operation.
This poses an additional challenge in the design of element-wise tunable \glspl{RIS} using the \gls{RA} method, since the bias lines will affect the performance of the radiating elements. 
This is commonly solved by biasing lines perpendicular to the electric-field polarization which is only possible for single, linearly polarized operation. Dual polarization designs require alternative solutions that account for the effect of the biasing lines when designing the radiating elements.
In the \gls{PA} method, this is not an issue due to the separation of the phase shifters and the radiating elements in different layers.


\subsection{Bandwidth}

\glspl{RIS} can be designed to support a single service provider (i.e., limited part of the band); an entire frequency band (e.g., 5G \SIrange{26.5}{29.5}{\giga\hertz} band), or even several bands, for example, the \SI{28}{\giga\hertz} and \SI{60}{\giga\hertz}. 
For high bandwidth systems, operating in a single band, the \gls{PA} implementation is more suitable compared to the \gls{RA}. However, \gls{RA} implementation is more suitable for multi-band operations due to the simplicity of their design. However, this comes at the cost of higher response time. Moreover, the phase-shift vs. voltage control curve of \gls{NLC}-\glspl{RIS} (see Fig.~\ref{fig:LCmolres}) changes across different frequencies within the band. Therefore, wideband phase-shift designs that exploit the frequency-dependent characteristics of the \gls{NLC}-\glspl{RIS} have to be developed.
\section{Conclusion}
\label{sc:conclusion}

This paper presented \gls{NLC} as an enabler technology for building extremely large \glspl{RIS} with continuous tuning capability, low power consumption, frequency scalability, and cost-efficient fabrication. However, these advantages come with the limitations such as slow response time and temperature dependencies. The basic physical principles of \gls{NLC} theory were reviewed and two important implementations of \gls{NLC}-\glspl{RIS} were introduced. Finally, \gls{NLC}-\glspl{RIS} were compared against competing technologies \gls{SC}- and \gls{MEMS}-\glspl{RIS} and several important open research problems on  \gls{NLC}-\glspl{RIS} were presented.

\ifCLASSOPTIONcaptionsoff
  \newpage
\fi

\bibliographystyle{IEEEtran}

\bibliography{literature}

\section{Acknowledgements}
This research was partly funded by the Deutsche Forschungsgemeinschaft (DFG, German Research Foundation) – Project-ID 287022738 – TRR 196 MARIE within project C09, by the mm-Cell project and the Collaborative Research Center 1053 MAKI, and by the LOEWE initiative (Hesse, Germany) within the emergenCITY center.

\section*{Biography}
\noindent 
\textbf{Alejandro Jim\'enez-S\'aez} is a Research Group Leader at the Institute of Microwave Engineering and Photonics (IMP) at the Technical University of Darmstadt (TUDa), Darmstadt, Germany. His research interests include RF components based on artificial and functional materials.

\vspace{0.3cm}
\noindent 
\textbf{Arash Asadi} (Senior Member, IEEE) is a Research Group Leader at TUDa, where he leads the Wireless Communication and Sensing Lab (WISE). His research interests include wireless communication and sensing and its application in beyond-5G/6G networks. 

\vspace{0.3cm}
\noindent 
\textbf{Robin Neuder} is a Ph.D. student at IMP, TUDa, Darmstadt, Germany.

\vspace{0.3cm}
\noindent 
\textbf{Mohamadreza Delbari} is a Ph.D. student at the Resilient Communication Systems (RCS) Lab at TUDa, Darmstadt, Germany.

\vspace{0.3cm}
\noindent 
\textbf{Vahid Jamali} is an Assistant Professor leading the RCS Lab at TUDa, Darmstadt, Germany. His research interests include wireless and molecular communications.

\end{document}